**Title :** Pricing cryptocurrencies : Modelling the ETHBTC spot-quotient variation as a diffusion process.

**Author :** Sidharth Mallik

**Abstract :** This research proposes a model for the intraday variation between the ETHBTC spot and the quotient of ETHUSDT and BTCUSDT traded on Binance. Under conditions of no-arbitrage, perfect accuracy and no microstructure effects, the variation must be equal to its theoretically computed value of 0. We conduct our research on 4 years of data. We find that the variation is not constantly 0. The variation shows a fluctuating behaviour on either side of 0. Furthermore, the deviations tend to be larger in the first year than the rest of the years. We test the sample for the nature of diffusion where we find evidence of mean-reversion. We model the variation using an Ornstein-Uhlenbeck process. A maximum likelihood estimation procedure is used. From the accuracy of the sampling distribution of the parameters obtained, we conclude that the variation may be accurately modelled as an Ornstein-Uhlenbeck process. From the parameters obtained, the long-term mean is shown to have a negative sign and differs from the theoretical value of 0 at 1e-05 precision. We take note of the results in light of efficiency of the markets to price publicly known information.

**Key messages :**
- The ETHBTC Spot-quotient variation indicates to be mean-reverting.
- The variation may be accurately modelled as an Ornstein-Uhlenbeck process.
- The result obtained suggests the existence of market inaccuracies in cryptocurrencies.

**Main Text :**

**Introduction :** Modelling financial series as diffusion processes with continuous-time characterization is well established like in the case of Vasicek(1977), Merton (1973) and Black and Scholes (1973) and many others. Frequently used models for financial series are the Geometric Brownian Motion which is expressed as the stochastic differential equation :

$$dS_t = \mu S_t dt + \sigma S_t dW_t$$

and the one that we explore later in more details, the Ornstein-Uhlenbeck process (OU process) expressed as the stochastic differential equation :

$$dv_t = \alpha(\mu - v_t)dt + \sigma dW_t$$

We here explore the use of diffusion models in the case of cryptocurrencies. These are relatively new instruments in the trading world with comparatively less amount of research and data available. The Geometric Brownian Motion has recently been tried with cryptocurrencies in Dipple *et al* (2020) and Cretarola and Figa-Talamanca (2018).

We consider a variation calculation in cryptocurrencies that involves three different instruments, as has been detailed later. We conduct preliminary examination of the sample. We then apply the Dickey-Fuller test based on Dickey and Fuller (1979) and Dickey (1976). As shown in Kim and Park (2014), this test may be used to check for mean-reversion for diffusion processes. We then model the variation based on the OU process, based on Uhlenbeck and Ornstein (1930). A maximum likelihood estimation process is used. We are able to obtain exact expressions for the parameters making this method a convenient way to estimate parameters in the given context. Goodness of fit is analyzed from a Monte Carlo simulated sampling distribution. From the obtained estimates of the parameters, we demonstrate that the variation shows deviation from its theoretical value by a small but non-negligible amount with a negative sign.

From the results obtained we make a note on the efficiency of the market in terms of pricing information. We refer to Fama (1970) and Fama (1991) for the hypothesis proposed for market efficiency. According to the hypothesis, three different forms of efficiency are considered based on the relevant information set. First, weak form, in which the information set is just the historical prices. Then semi-strong form, in which the concern is whether prices efficiently adjust to other information, that is obviously publicly available (e.g., announcements of annual earnings, stock splits, etc.). Finally, strong form concerned with whether given investors or groups have monopolistic access to any information relevant for price formation. Research on lines of market efficiency has been conducted for cryptocurrency markets more recently by Tran and Leirvik (2020) and by Lopez-Martin *et al* (2021). Tran and Leirvik (2020) conclude that the efficiency in cryptocurrency markets varies with time and individual currencies. They mention that one reason for risks and inefficiencies in these markets can be that the markets have been difficult to trade in and hence liquidity has been low compared to other markets. The ease of trading one cryptocurrency can be significantly different from the ease of trading another such currency, thus the liquidity in various such currencies varies substantially. They also find that markets for cryptocurrencies are improving at an exceptional pace, with volume improving and becoming less volatile. Lopez-Martin *et al* (2021) report a similar behaviour noting that while cryptocurrency markets have been inefficient in the recent past, their efficiency is improving with an increase in liquidity. While our research does not explicitly conduct tests for market efficiency, the results we obtain may be considered an observation towards the efficiency of the markets under study. More specifically, we focus on the semi-strong form of the hypothesis on market efficiency.

**Spot-quotient variation :** Cryptocurrencies are relatively newer trading instruments that trade on several exchanges across the world. Binance is one of the most prominent cryptocurrency exchanges in the world being regularly in the top 10 ranked cryptocurrency exchanges on liquidity measures. Instruments in binance are usually of two types, either denominated in USA Dollars (USDT) or denominated in another cryptocurrency. For example, the instrument BTCUSDT refers to the cryptocurrency called Bitcoin (BTC) denominated in USA Dollars (USDT). Similarly ETHUSDT refers to the cryptocurrency called Ethereum (ETH) denominated in USA Dollars (USDT). For the other type of cryptocurrency, a well traded example is ETHBTC which is the cryptocurrency Ethereum (ETH) denominated in another cryptocurrency, which, in this case is Bitcoin (BTC). Such an arrangement creates the possibility of the spot-quotient variation.

In the spot-quotient variation, the spot leg comes from cryptocurrencies denominated in other cryptocurrencies whereas the quotient leg comes from dividing one cryptocurrency with another, both of which are denominated in USA Dollars. As an example, when ETHBTC is the spot leg of the variation, the quotient leg comes from the quotient with ETHUSDT in the numerator and BTCUSDT in the denominator. Finally we take logarithms of each of the legs before calculating the difference. So the i-th value of the variation is given as

$$\ln(ETHBTC[i]) - \ln\left(\frac{ETHUSDT[i]}{BTCUSDT[i]}\right) \qquad (1)$$

The two legs are representative of the same value and hence the difference, in theory, is 0. We investigate the behaviour of such a variation.

**Data description :** The data has been obtained directly from the Binance exchange. We research data from September $1^{st}$, 2017 to August $31^{st}$, 2021. The data is sampled at 1 minute time intervals with the ending price at the interval taken for construction of the variation. The size of the sample comes out to be over 2 million for the period considered.

**Procedure :** The research is conducted in following phases :
- We conduct preliminary examination of the sample
- We perform the Dickey-Fuller test
- On getting confirmation from the tests, we model the variation using the OU process
- We perform inference on the parameters of the model
- We conclude the research using the results

**Preliminary examination of the variation :** The table of percentiles clearly show that the variation is not constantly 0. It has a tendency to deviate on either side of 0.

| Percentile | Value |
|---|---|
| 0 | -0.084171 |
| 25 | -0.000161 |
| 50 | -9.312451e-07 |
| 75 | 0.000155 |
| 100 | 0.092048 |

Table 1 : Table of percentiles for the variations

We further break the data down yearwise and construct the table of percentiles for each of the years separately. Year 1 tends to have greater larger deviations than the other years.

| Percentile → | 0 | 25 | 50 | 75 | 100 |
|---|---|---|---|---|---|
| Year 1 | -0.084171 | -0.000795 | -4.186517e-05 | 0.000688 | 0.092048 |
| Year 2 | -0.025374 | -0.000207 | 3.573059e-06 | 0.000214 | 0.011835 |
| Year 3 | -0.037243 | -9.982146e-05 | 2.214119e-06 | 0.000104 | 0.0149723 |
| Year 4 | -0.024111 | -7.368830e-05 | -6.181590e-07 | 7.340599e-05 | 0.020280 |

Table 2 : Yearwise table of percentiles for the variations

We further calculate the interquartile ranges for each of the years. This is based on the difference between the 75 percentile and the 25 percentile values.

| Year number | Interquartile Range |
|---|---|
| Year 1 | 0.001483 |
| Year 2 | 0.000421 |
| Year 3 | 0.000204 |
| Year 4 | 0.000147 |

Table 3 : Yearwise interquartile range for the variations

The interquartile range clearly shows a decreasing trend along the years.

**The Dickey-Fuller test :** As in Dickey and Fuller (1979), the three models considered are :
$$Y_t = \rho Y_{(t-1)} + e_t, where\ t = 1, 2, \ldots$$
$$Y_t = \mu + \rho Y_{(t-1)} + e_t, where\ t = 1, 2, \ldots$$
$$Y_t = \mu + \beta t + \rho Y_{(t-1)} + e_t, where\ t = 1, 2, \ldots$$

We perform a change in coefficients to obtain the following forms,
$$Y_t - Y_{(t-1)} = \delta Y_{(t-1)} + e_t, where\ t = 1, 2, \ldots \quad (a)$$
$$Y_t - Y_{(t-1)} = \mu + \delta Y_{(t-1)} + e_t, where\ t = 1, 2, \ldots \quad (b)$$
$$Y_t - Y_{(t-1)} = \mu + \beta t + \delta Y_{(t-1)} + e_t, where\ t = 1, 2, \ldots \quad (c)$$

Tests are conducted for the coefficient of $Y_{t-1}$ and based on the results the nature of the time series is characterized. Specifically a rejection of the null hypotheses on all three models is considered an indicator of mean-reversion. The table values for comparison are taken from Dickey (1976).

**Result of Dickey-Fuller test :** We test for the three different hypothesis tests mentioned above. We considered a 1% significance level. As shown in the table, at the given significance level, the null hypothesis is rejected for all the three scenarios. This indicates a mean-reverting diffusion process

| Model | Acceptance or Rejection of null hypothesis |
| --- | --- |
| Model (a) | Rejected |
| Model (b) | Rejected |
| Model (c) | Rejected |

Table 4 : Results of the Dickey-Fuller test

**Characterization of normal distribution for the spot-quotient variation :** Under the assumption of a Geometric Brownian Motion, the price of BTCUSDT would follow a lognormal distribution. So would the price of ETHUSDT. We further assume that the prices follow a joint lognormal distribution. Lognormal distributions have the property that their quotient is again a lognormal distribution. This may be easily demonstrated as follows.

Let $X_1 \sim Normal(m_1, s_1), X_2 \sim Normal(m_2, s_2)$ where, 'm' and 's' represent the corresponding means and standard deviations.
We take, $Y_1 = \exp(X_1), Y_2 = \exp(X_2)$
Then, $Y_1 \sim Lognormal(m_1, s_1), Y_2 \sim Lognormal(m_2, s_2)$ where, 'm' and 's' represent the corresponding means and standard deviations of the underlying normal distributions.
We consider $\frac{Y_1}{Y_2} = \frac{\exp(X_1)}{\exp(X_2)} = \exp(X_1 - X_2)$

Now, difference of two normally distributed variables is again a normally distributed variable.
Since, $(X_1 - X_2) \sim Normal(m_{12}, s_{12})$
Therefore, $\dfrac{Y_1}{Y_2} \sim Lognormal(m_{12}, s_{12})$

Using the above derivation, each of the legs of the spot-quotient variation may be considered normally distributed. Further, the two legs are representing the same value, hence we assume that the legs are jointly normally distributed. Since the variation in (1) is simply a difference between the legs, it may be characterized as normally distributed.

**The OU process :** As mentioned earlier the OU process has been used to model other financial time series successfully. The representation is as :
$$dv_t = \alpha(\mu - v_t)dt + \sigma dW_t$$
$dW_t$ represents Brownian Motion. μ represents the long-term mean, σ represents the variance parameter and α represents the elasticity parameter. The series has the tendency to revert back to its long-term mean. The process is continuous and has a stationary distribution. The first term is responsible for the reversion to the long-term mean and its magnitude increases with the magnitude of the deviation from the long-term mean. The reversion intensity is amplified by the elasticity parameter α. The dependence on Brownian Motion implies that the process has normal increments. We use this property in parameter estimation. The variance parameter σ amplifies the amplitude of the Brownian Motion. The conditional expectation and variance of the process given the current level are
$$E_t v_s = \mu + (v_t - \mu)\exp(-\alpha(s-t)) \, where \, (t \leq s)$$
$$Var_t v_s = \dfrac{\sigma^2}{2\alpha}(1 - \exp(-2\alpha(s-t))) \, where \, (t \leq s)$$

**Maximum Likelihood Estimation of the OU process :** Maximum Likelihood Estimation is an established method of parameter estimation in situations where assumptions regarding underlying distributions can be made. Since we are working with an OU process, as mentioned before, the process has normal increments. We make use of this fact to construct the likelihood function. The conditional probability density is :
$$f(v_s | v_t; \mu, \alpha, \hat{\sigma}) = \dfrac{1}{\hat{\sigma}\sqrt{2\pi}} \exp\left(\dfrac{-(v_s - \mu - (v_t - \mu)\exp(-\alpha(s-t)))^2}{2\hat{\sigma}^2}\right)$$
$$\text{where,} \quad \hat{\sigma}^2 = \sigma^2 \dfrac{1 - \exp(-2\alpha(s-t))}{2\alpha}$$

From the conditional probability density, we obtain the log-likelihood function as :
$$\Lambda(\mu, \alpha, \hat{\sigma}) = \sum_{i=1}^{n} \ln(f(v_i | v_{(i-1)}); \mu, \alpha, \hat{\sigma})$$
$$= \dfrac{-n}{2}\ln(2\pi) - n\ln(\hat{\sigma}) - \dfrac{1}{2\hat{\sigma}^2} \sum_{i=1}^{n} [v_i - \mu - (v_{(i-1)} - \mu)\exp(-\alpha)]^2$$

Setting,

$$\dfrac{\delta \Lambda(\mu, \alpha, \hat{\sigma})}{\delta \mu} = 0$$
$$\dfrac{\delta \Lambda(\mu, \alpha, \hat{\sigma})}{\delta \alpha} = 0$$

$$\frac{\delta \Lambda(\mu, \alpha, \hat{\sigma})}{\delta \hat{\sigma}} = 0$$

and solving for the resulting equations, we arrive at the exact solutions for the parameters.
Solution of the parameters :
To express the solutions we use the following notations :

$$S_0 = \sum_{i=1}^{n} v_{(i-1)} \quad S_1 = \sum_{i=1}^{n} v_{(i)} \quad S_{00} = \sum_{i=1}^{n} v_{(i-1)}^2 \quad S_{01} = \sum_{i=1}^{n} v_{(i-1)} v_{(i)} \quad S_{11} = \sum_{i=1}^{n} v_{(i)}^2$$

We obtain,

$$\mu = \frac{S_1 S_{00} - S_0 S_{01}}{n(S_{00} - S_{01}) - (S_0^2 - S_{01})}$$

$$\alpha = -\ln\left(\frac{S_{01} - \mu S_0 - \mu S_1 + n\mu^2}{S_{00} - 2\mu S_0 + n\mu^2}\right)$$

$$\hat{\sigma}^2 = \frac{1}{n}[S_{11} - 2\omega S_{01} + \omega^2 S_{00} - 2\mu(1-\omega)(S_1 - \omega S_0) + n\mu^2(1-\omega)^2]$$

where, $\omega = \exp(-\alpha)$
and, $\sigma^2 = \hat{\sigma}^2 \frac{2\alpha}{1-\omega^2}$

**Result of MLE parameter estimation of OU process :** We obtain the values for the parameters of the OU process as displayed in the following table :

| Parameter | Value |
|---|---|
| $\hat{\alpha}$ | 0.845728 |
| $\hat{\mu}$ | -2.424382e-05 |
| $\hat{\sigma}$ | 0.001703 |

Table 5 : Results of the MLE estimation of OU process

Simulating a sampling distribution for the parameters obtained : We estimate the sampling distribution of the parameters by Monte Carlo sampling. We report 90% confidence intervals for each of the parameters with the number of samples as 1000. The results are displayed in the following table :

| Parameter | Lower Bound | Upper Bound |
|---|---|---|
| $\hat{\alpha}$ | 0.842325 | 0.847948 |
| $\hat{\mu}$ | -2.801373e-05 | -2.207785e-05 |
| $\hat{\sigma}$ | 0.001700 | 0.001705 |

Table 6 : Confidence intervals of the parameters estimated

**Inference on parameters :** The sampling distribution indicates that the parameters have been estimated accurately. The estimate of the long-term mean $\mu$ differs from 0 at 1e-05 precision. Its confidence interval indicates a negative sign. The estimate of the elasticity parameter $\alpha$ is strictly positive across the confidence interval as required for the OU process. Based on the estimate of all the three parameters and the obtained sampling distribution, we infer that the variations may be accurately modelled using an OU process.

**Conclusions :**
1. The variations have been successfully shown to be of mean-reverting nature with an accurately fit diffusion process. The significantly different value of the long-term mean from its theoretical value of 0 at a precision of 1e-05 may be considered small but still significant.
2. Since the two legs of the variation represent the same value, a significant deviation from 0 indicates the existence of inaccuracies in the market. Seeing this in light of the semi-strong form of hypothesis of market efficiency, it may be observed that all public information has not been completely incorporated in the prices leading to the existence of the non-zero variation.
3. As other studies have also pointed out, the efficiency of the market is increasing with time. Based on the preliminary analysis of the variation, we observe a similar trend in the yearwise data. The deviation of the variation from 0 has been reducing with time.

**Acknowledgements :**
1. We acknowledge the use of facilities provided by the University of Hull. This includes access to journals for matter of reference.
2. We acknowledge the use of data provided by Binance for our research.
3. We acknowledge the use of the Python programming language for our research.

**Declarations of Interest :** The author reports no conflict of interest. The author alone is responsible for the content and writing of the paper.

**References :**
**Black, F., & Scholes, M. (1973),** The Pricing of Options and Corporate Liabilities. *Journal of Political Economy*, *81*(3), 637–654. http://www.jstor.org/stable/1831029
**Cretarola, A. And Figa-Talamanca, G. (2018)**, Modeling Bitcoin Price and Bubbles, Blockchain and Cryptocurrencies, Asma Salman and Muthanna G. Abdul Razzaq, IntechOpen, DOI: 10.5772/intechopen.79386. Available from: https://www.intechopen.com/chapters/62627
**Dickey, D. A. (1976)**, Estimation and hypothesis testing in nonstationary time series. *Retrospective Theses and Dissertations*. 6267. https://lib.dr.iastate.edu/rtd/6267
**Dickey, D. A., and Fuller, W. A. (1979)**, Distribution of the estimators for autoregressive time series with a unit root, Journal of the American Statistical Association, 74:366a, 427-431, DOI:10.1080/01621459.1979.10482531
**Dipple, S., Choudhary, A., Flamino, J., Szymanski, B. K. and Korniss, G. (2020)**, Using correlated stochastic differential equations to forecast cryptocurrency rates and social media activities. *Applied Network Science* 5, 17 (2020). https://doi.org/10.1007/s41109-020-00259-1
**Fama, E. F. (1970),** Efficient Capital Markets: A Review of Theory and Empirical Work. *The Journal of Finance*, *25*(2), 383–417. https://doi.org/10.2307/2325486

**Fama, E.F. (1991)**, Efficient capital markets: II. *The Journal of Finance.* https://doi.org/10.1111/j.1540-6261.1991.tb04636.x

**Kim, J., and Park, J. Y. (2014)**, Mean reversion and the unit root properties of diffusion models

**Lopez-Martin, C., Muela, S.B. and Arguedas, R. (2021)**, Efficiency in cryptocurrency markets: new evidence. *Eurasian Econ Rev* 11, 403–431 (2021). https://doi.org/10.1007/s40822-021-00182-5

**Merton, R. C. (1973),** Theory of Rational Option Pricing. *The Bell Journal of Economics and Management Science*, *4*(1), 141–183. https://doi.org/10.2307/3003143

**Tran, V. L. and Leirvik, T. (2020)**, Efficiency in the markets of crypto-currencies. Finance Research Letters, 35, https://doi.org/10.1016/j.frl.2019.101382

**Uhlenbeck, G. E., and Ornstein, L. S. (1930)**, On the theory of the Brownian Motion. American Physical Society, 36(5), 823-841. https://link.aps.org/doi/10.1103/PhysRev.36.823

**Vasicek, O. (1977)**, An equilibrium characterization of the term structure. Journal of Financial Economics, 5(2), 177-188. https://doi.org/10.1016/0304-405X(77)90016-2